\documentclass[
reprint,
amsmath,amssymb,
]{revtex4-2}

\usepackage{graphicx}
\usepackage{dcolumn}
\usepackage{bm}
\usepackage{color}
\usepackage{comment}
\usepackage{txfonts}
\usepackage{hyperref}
\hypersetup{
colorlinks=true,
hidelinks
}

\begin{document}

\title{Dynamic Scaling of Vorticity in Phase-separating Superfluid Mixtures}

\author{Ryuta Ito$^{1}$}
% \email{sa24426b@st.omu.ac.jp}
\author{Hiromitsu Takeuchi$^{1,2}$}
 \email{takeuchi@omu.ac.jp}
\affiliation{
$^1$ Department of Physics, Osaka Metropolitan University, 3-3-138 Sugimoto, Osaka 558-8585, Japan \\
$^2$ Department of Physics, Nambu Yoichiro Institute of Theoretical and Experimental Physics (NITEP), Osaka Metropolitan University, 3-3-138 Sugimoto, Osaka 558-8585, Japan
}

\date{\today}

%%%%%%%%%%%%%%%%%%%%%%%%%%%%%%%%%%%%%%%%%%%%%%%%%%%%%%%%%%%%%%%%%%%%%%%%%%%%%%%%%%%%%%%%%%%%%%%%

\begin{abstract}
Recently, it has been experimentally confirmed that non-equilibrium dynamics of phase separation in strongly ferromagnetic Bose-Einstein condensates of $^7$Li atoms obey the dynamic scaling law belonging to the binary-fluid universality class in the inertial hydrodynamic stage.
The current work theoretically and numerically studies the dynamic scaling law of structure factor of vorticity in a phase-separating binary superfluid mixture, equivalent to the $^7$Li condensates in a strong limit of quadratic Zeeman shift.
We found a dynamic scaling law for the structure factor based on our numerical observation that the peak of the energy spectrum from turbulence theory does not vary in time in the stage.
Similarly to freely decaying turbulence, a power-law hierarchy exists in the energy spectrum in our system,
and we proposed a decay law of the energy based on the dynamic scaling law
by introducing the microscopic, high-wavenumber cutoff.
\end{abstract}

\maketitle

The dynamic scaling law of phase ordering kinetics states empirically that, in the non-equilibrium time evolution of a phase transition accompanying spontaneous symmetry breaking, the spatial structure of the order parameter fields evolves in time while preserving statistical similarity~\cite{Bray1994}.
This empirical law of phase transitions is universally believed to hold for a variety of physical systems and has been confirmed numerically and experimentally in binary alloys~\cite{Gunton1990}, cell membranes~\cite{Camley2011}, and binary quantum fluids~\cite{Hoffer1986,Damle1996,HT2012,Kudo&Kawaguchi2013,Hofmann2014,Kudo&Kawaguchi2015,Blakie2016,Williamson2017,Karl2017,HT2018,Williamson2019,Fujimoto2020,Williamson2021,KAIST2024}
%\HT{[Takeuchi2010,Wiliamson,Wiliamson,]}
as well as classical fluids~\cite{Burstyn1983,Furukawa1985,Farrell1989,Alexander1993,Xu2005,Ahmad2012,Tanaka2015,Chandrodoy2023}.
%\HT{[motto-aru-hazu]}.
When discrete symmetry breaking occurs,
such as in the ferromagnetic phase transition of the two-dimensional Ising model,
the structure factor or correlation function of longitudinal magnetization is described by a universal function that is independent of time by rescaling in an appropriate manner with the domain characteristic size $l$ that obeys the power-law growth with time: $l(t)\propto t^{1/z}$~\cite{Bray1994}.
The domain walls or interfaces between spatial domains with different ordered states have a macroscopic geometry that universally follows the percolation criticality~\cite{Leticia2007,Barros2009,HT2015}.
Recently, two-dimensional phase separation dynamics of ferromagnetic superfluid mixture was realized for the first time in the experiment of strongly ferromagnetic Bose-Einstein condensates (BECs) of $^7$Li atoms~\cite{KAIST2024},
where the dynamic scaling law holds for the structure factor of longitudinal magnetization in the inertial hydrodynamic stage with the dynamic exponent $1/z=2/3$.\par

In the inertial hydrodynamic stage for ordinary viscous or classical fluid mixtures, 
the viscous term is negligible compared with the inertial term in the Navier-Stokes equation~\cite{Bray1994}.
In this sense, it may be reasonable that both classical fluid and non-viscous superfluid mixtures have statistically universal properties in this stage.
Although fluid flows significantly influence the mesoscopic structure in this stage,
it is not at all clear how the mesoscopic structure of flow or turbulence in the presence of domain walls relates to the dynamic scaling law of the mixture.
The complex motion of interfaces in the phase separation process is thought to be coupled with the flow,
forming vortex sheets along domain walls and thus a sort of vortex sheet turbulence that may play a crucial role to connect the mesoscopic distribution of longitudinal magnetization to that of vorticity.
Indeed, in the late stage of the phase separation process in superfluid mixtures,
quantum Kelvin-Helmholtz instability (KHI)~\cite{HT2010,Kokubo2021,HT2022,Huh2024} at domain walls generates quantum vortices as magnetic skyrmions,
which can cause anomalous statistical behavior for small-scale domains differently from those in flow-free Ising systems~\cite{HT2018}.
Even in the interface dynamics in classical fluid mixture such as viscous fingering, fluid flow plays an important role in the formation of characteristic interface patterns and influence its fractal structure~\cite{Nittmann1985,Homsy1987}.
While extensive researches have been conducted on the structural and geometric properties for different mixture systems~\cite{Gunton1990,Camley2011,Hoffer1986,Damle1996,HT2012,Kudo&Kawaguchi2013,Hofmann2014,Kudo&Kawaguchi2015,Blakie2016,Williamson2017,Karl2017,HT2018,Williamson2019,Fujimoto2020,Williamson2021,KAIST2024,Burstyn1983,Furukawa1985,Farrell1989,Alexander1993,Xu2005,Ahmad2012,Tanaka2015,Chandrodoy2023,Leticia2007,Barros2009,HT2015}, studies focusing on their hydrodynamic aspects remain scarce.
\par

As a stepping stone to challenging these issues, this letter investigates the non-equilibrium dynamics of the structure factor of vorticity in non-viscous superfluid mixtures by numerically simulating phase separation in the Gross-Pitaevskii (GP) model~\cite{Pethick&Smith2008}.
We numerically establish a dynamic scaling law for the vorticity structure factor by relating the energy spectrum from turbulence theory, $E(k)$ with wavenumber $k$~\cite{Batchelor1948,Comte-Bellot1966,Kida1981}.
By applying an analysis that extends the dynamic hierarchy hypothesis~\cite{HT2018} to the energy spectrum,
we show that a decay law similar to that of freely decaying turbulence appears in our system.
The energy spectrum of our superfluid mixture asymptotically approaches the power law $E(k)\propto k^{-1}$ of an uncorrelated vortex distribution for high wavenumbers.\par

\begin{figure*}[htbp]
    \centering
    \includegraphics[width=1.0\linewidth]{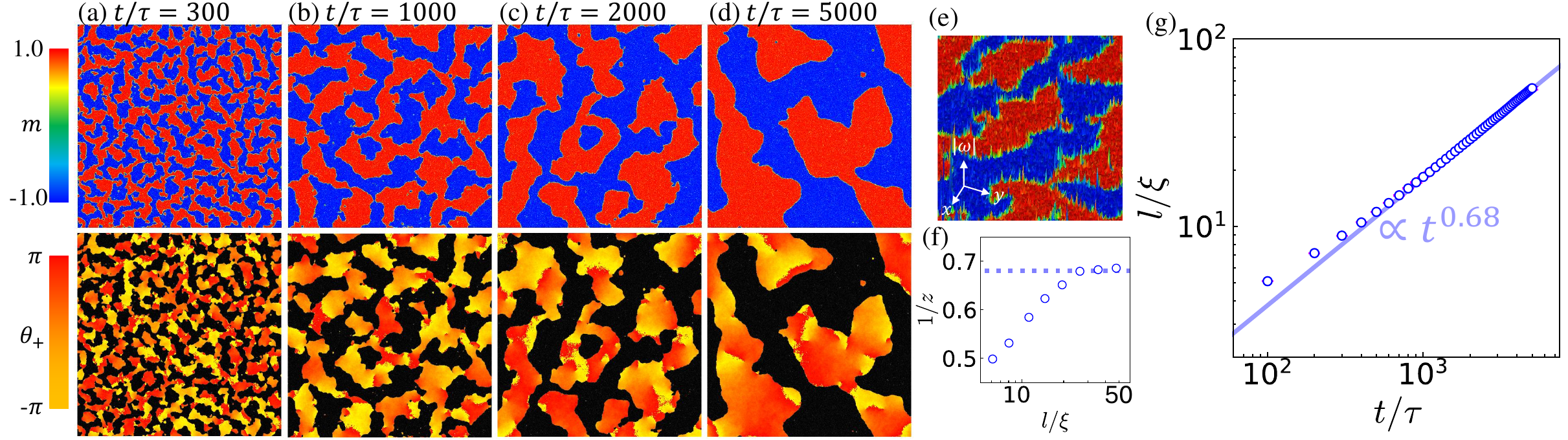}
    \caption{
    (Color online) (a-d) Snapshots of phase-separating superfluid mixtures with the system size $512\xi$ on each side. (Top) The polarization $m(\bm{r})=(n_+(\bm{r})-n_-(\bm{r}))/(n_+(\bm{r})+n_-(\bm{r}))$ from $t/\tau=300$ to $5000$ with $\tau=\hbar/\mu$ is shown. The $+$ ($-$) domains correspond to red (blue) regions. (Bottom) The phase of the $|+\rangle$ component, $\theta_+=\arg (\psi_+)$, is shown. A vortex corresponds to the endpoint of a branch cut. (e) An enlarged picture of the system at $t/\tau=300$. The height indicates absolute value of vorticity $\omega=(\nabla\times \bm{v})_z$. (f) The time dependence of dynamic exponent $z$. The dotted line indicates $1/z=0.68$. (g) The time evolution of the characteristic length scale. The late-time dynamics obeys the power law with $l\propto t^{0.68}$.
    }
    \label{fig:1}
\end{figure*}

{\it Dynamic Hierarchy Hypothesis.}
In order to formulate the dynamic scaling law of the structure factor of vorticity, we first define the flow velocity and vorticity in our superfluid mixtures and discuss their relation with the characteristic length. Following the convention in the literature of multi-component classical and quantum fluids~\cite{Landau1987,Konso1991,Kawaguchi2012,Vollhardt2013,Tsubota2013}, we define the total density $\rho$ and the flow velocity $\bm{v}$ of the fluid mixture as follows
\begin{align} \label{velocity}
    \rho=\sum_j\rho_j,\hspace{2mm}\bm{v} = \frac{1}{\rho}\sum_j\rho_j\bm{v}_j
\end{align}
where $\rho_j$ and $\bm{v}_j$ are the mass density and the velocity of $j$-th fluid component, respectively. Considering two-dimensional flows, the vorticity can be written as
$\bm{\omega}=\nabla\times\bm{v}=(0,0,\omega)^{{\rm T}}$. The vorticity structure factor is given by
\begin{align} \label{s_v}
   S_{\rm v}(k,t) = \langle |\hat{\omega}(\bm{k},t)|^2 \rangle,
\end{align}
where $\hat{\omega}(\bm{k},t)=\int \omega(\bm{r},t)e^{-i\bm{k}\cdot\bm{r}}d^2x=\hat{\omega}^{*}(\bm{-k},t)$ is the Fourier transform of vorticity and the bracket $\langle\cdots\rangle$ denotes the angular average with respect to the wavenumber vector $\bm{k}$.\par

In the late stage of the phase separation process, the characteristic length $l$, or the characteristic domain size, is a fundamental quantity to rescale the non-equilibrium dynamics.
% and obeys a power law $l(t) \propto t^{1/z}$, where $z$ is called the dynamic exponent.
Our main goal is to find a scale exponent $\alpha$ for the vorticity structure factor in the following form~\cite{form}
\begin{align} \label{Scaling_function_Sv}
    \tilde{S}_{\rm v}(\tilde{k}) \propto S_{\rm v}(k,t)l(t)^{\alpha}
\end{align}
with the rescaled wavenumber $\tilde{k}=kl$. The rescaled structure factor $\tilde{S}_{\rm v}$ is called the universal function of vorticity structure factor and is explicitly independent of time. This requirement is due to the dynamic scaling hypothesis that all physical quantities maintain their statistical similarity throughout time evolution.

To study the flow structure, it is useful to consider the energy spectrum from turbulence theory~\cite{Frisch1995,Davidson2004}.
It is defined with $S_{\rm v}$ as
$E = L^2S_{\rm v}/(4\pi k)$ with the system size $L$. Then one obtains the universal function of energy spectrum
\begin{align} \label{Scaling_function_E}
    \tilde{E}(\tilde{k}) \propto \frac{\tilde{S}_{\rm v}(\tilde{k})}{\tilde{k}}l(t)^{1-\alpha}.
\end{align}

According to Ref.~\cite{HT2018}, it is important to explicitly address the hierarchy with respect to the length scales. The difference in statistical behavior distinguishes between the macroscopic regime with scales larger than $l$ and the microscopic one smaller than it; the former and latter regimes are defined as $l/L<\tilde{k}<1$ and $1<\tilde{k}<l/\xi$, respectively.
Here, we introduced the microscopic scale $\xi$ as the thickness of domain wall between the two fluid components. Accordingly, we assume asymptotic forms for the universal functions with the length hierarchy as
\begin{align} \label{Asymp_scaling}
        \tilde{S}_{\rm v} \propto
    \begin{cases}
        \tilde{k}^{1+a} \\
        \tilde{k}^{1+b} 
    \end{cases} \hspace{-2mm}
    \text{and} \hspace{2mm}
        \tilde{E} \propto
    \begin{cases}
        \tilde{k}^{a} \hspace{2mm} \text{for} \hspace{2mm} l/L \ll \tilde{k} \ll 1, \\
        \tilde{k}^{b} \hspace{2mm} \text{for} \hspace{2mm} 1 \ll \tilde{k} \ll l/\xi.
    \end{cases}
\end{align}
This is the fundamental assumption basic to our theory.

{\it Phase Separation of a Superfluid Mixture.}
To numerically examine our theory for the segregation dynamics of a superfluid mixture, we introduce the GP model for multi-component BECs at absolute zero~\cite{Pethick&Smith2008}. Here, we consider a system of spin-1 BECs in the ferromagnetic phase without linear Zeeman shift. In this phase, the $|0\rangle$ Zeeman component is negligible under a large negative quadratic Zeeman shift~\cite{HT2022}. In this region, the system reduces to a two-component BEC, and the GP Lagrangian in two dimensions is written as
\begin{align}
    \mathcal{L} = \int d^2x \left[\mathcal{P}_{+}+\mathcal{P}_{-}-g_{+-}|\psi_{+}|^2|\psi_{-}|^2 \right].
\end{align}
Here, we used
\begin{align}
    \mathcal{P}_{j} = i\hbar \psi _{j}^{*}\frac{\partial \psi _{j}}{\partial t} -\frac{\hbar ^{2}}{2m}|\nabla \psi _{j} |^{2} +\mu|\psi_j|^2 -\frac{g_{jj}}{2} |\psi _{j} |^{4}
\end{align}
with atomic mass $m$, the chemical potential $\mu$, the coupling constants $g_{ij}$ and the macroscopic wave function $\psi_j$ of the $|j\rangle$ component with $i,\,j=+,\,-$. The segregation takes place if the coupling constants satisfy the condition $g_{+-}>g$ with $g_{++}=g_{--}\equiv g>0$. We choose $g_{+-}/g=3$ which has been realized for $^7$Li atomic gases~\cite{KAIST2020,KAIST2024}. Then we have the exact solution of a flat domain wall between the $+$ and $-$ domains~\cite{Indekeu2015}.
The cross-section profile of the longitudinal magnetization for the solution is proportional to $\tanh(x/\xi)$,
where the flat wall is perpendicular to the $x$ direction and $\xi=\hbar/\sqrt{m\mu}$ is the wall thickness, called the healing length.\par

The numerical scheme is the same as in previous studies~\cite{HT2015,HT2018}. Our simulations start from the initial state with a homogeneous state $|\psi_+|^2=|\psi_-|^2=\bar{n}/2\,(={\rm const.})$ with the bulk density $\bar{n}\approx \mu/g$ after phase separation. We add small random seeds made by a white noise to the initial state to trigger the dynamic instability~\cite{num}.

We use the mean distance between domain walls as the characteristic length which is defined by $l=L^2/R$ with the total length of the domain walls $R$ and the system size $L$~\cite{length}.
Figure~\ref{fig:1}(g) shows the time evolution of the characteristic length. 
We find $1/z=0.68(1)$ consistent with $1/z=2/3$ in the inertial hydrodynamic stage, observed in the numerical and experimental studies~\cite{Kudo&Kawaguchi2013,Hofmann2014,Blakie2016,KAIST2024}. Note that the dynamic exponent changes over time and it takes the above value in the late stage for $l/\xi\gtrsim30$ [Fig.~\ref{fig:1}(f)].
It has been also pointed out that the dynamic exponent $z$ may depend on time in binary classical fluids~\cite{Farrell1989}.
%Previous studies in quantum fluid systems also support this numerically and experimentally~\cite{Hofmann2014,Blakie2016,KAIST2024}.

The relative velocity between different fluid components generates vorticity {\it charged} on domain walls, leading to the formation of vortex sheets [Fig.~\ref{fig:1}(e)]~\cite{vortex}.
Quantum KHI occurs at a domain wall and vortices or skyrmions are generated~\cite{HT2018,HT2022}.
These vortices carry vorticity away from the domain walls~\cite{HT2010}.
Unlike quantum vortices in single-component superfluids, vorticity in this system may be continuously distributed along domain walls.
The fractal geometry of domain wall due to the percolation criticality makes the vorticity distribution complicated.\par

\begin{figure}[htb]
    \centering
    \includegraphics[width=1.0\linewidth]{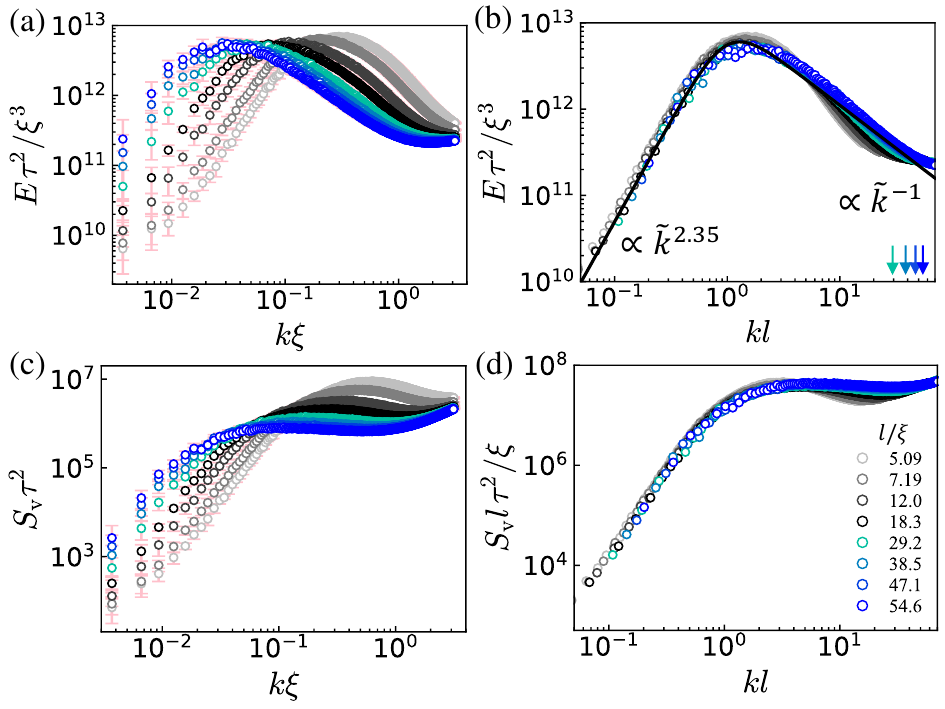}
    \caption{
    (Color online) The plots of the energy spectrum (a) and the vorticity structure factor (c), and the rescaled plots of the energy spectrum (b) and of the vorticity structure factor (d). The black line in (b) indicates our ansatz for the universal function of the energy spectrum [Eq.~\eqref{ansatz}]. The arrows in (b) indicate the upper cutoff $l/\xi$ of the integration. Angular average is taken only for the structure factor, and ensemble averages over 10 realizations are shown in all plots.
    }
    \label{scaling}
\end{figure}

{\it Dynamic Scaling Plots.}
The vorticity structure factor and energy spectrum are computed for different times
(Fig.~\ref{scaling}).
We took an ensemble average over 10 realizations with the random initial seeds after performing the angular average with respect to the wavenumber vector $\bm{k}$.
Here we explore the dynamic scaling law for the vorticity structure factor and energy spectrum.\par

First, let us evaluate the dynamic scaling law from the energy spectrum $E(k,t)$ that exhibits a simple time-dependence.
% The first operation to obtain a dynamic scaling plot is to rescale the wavenumber $k$ by $l$.
In Fig.~\ref{scaling}(a), we can see that 
the maximum value of $E(k, t)$ remains nearly constant in the inertial hydrodynamic stage ($l/\xi \gtrsim 30$).
Accordingly, when the energy spectrum is plotted as a function of $\tilde{k}$,
the plots collapse onto a single, time-independent function for $l/\xi \gtrsim 30$ [Fig.~\ref{scaling}(b)].
This heuristic evaluation implies that $\alpha=1$ in Eq.~(\ref{Scaling_function_E}).
In fact, using Eq.~(\ref{Scaling_function_Sv}) with $\alpha=1$,
we succeed in {\it collapsing} the vorticity structure factor as shown in Fig.~\ref{scaling}(d).\par

In the early stage ($l/\xi\lesssim30$), the rescaled plots of the energy spectrum do not coincide for the microscopic regime ($1< \tilde{k} < l/\xi$),
although a power-law behavior appears clearly in the macroscopic regime ($l/L<\tilde{k}<1$).
This behavior is consistent with the previous work~\cite{HT2018},
where the universal anomaly in the microscopic region of the domain area distribution only emerges late in the time evolution.\par

\begin{figure}[htb]
    \centering
    \includegraphics[width=1.0\linewidth]{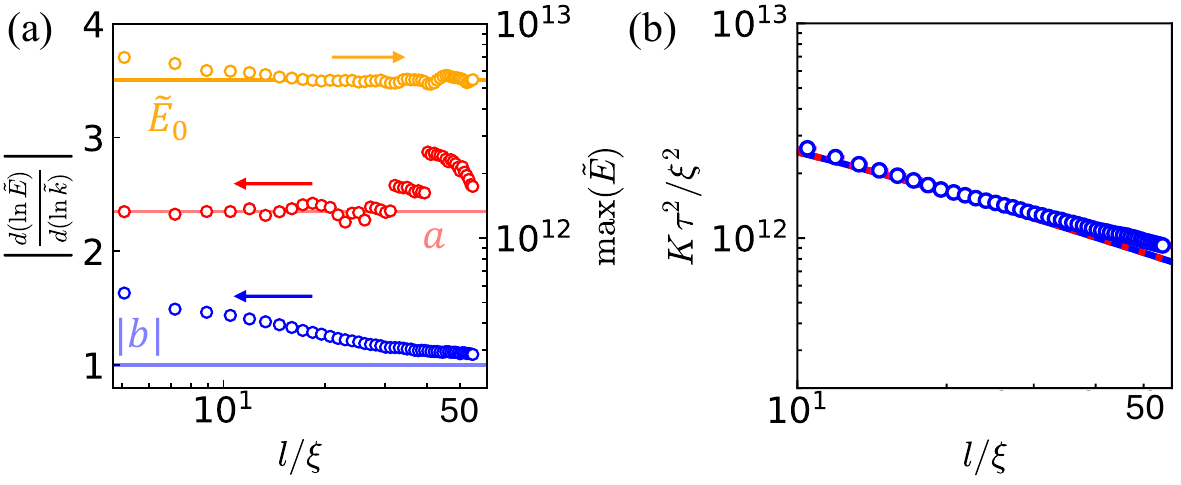}
    \caption{
    (a) (Color online) The time evolution of the maximum value of the energy spectrum $\max(\tilde{E})$ and the exponent $|d\ln\tilde{E}/d\ln\tilde{k}|$ in the macroscopic regime and in the microscopic regime, from top to bottom. Each line indicates $\tilde{E}_0=5.46\times10^{12}$, $a=2.35$ and $|b|=1$, from top to bottom.
    (b) The decay law of the vortex energy obtained numerically and theoretically. The blue line and the dotted red line indicates a decay law with Eqs.~\eqref{F_-1} and~\eqref{decay_asym}, respectively.
    }
    \label{fluid_energy}
\end{figure}

{\it Energy Decay Law.}
To gain more insight into the relationship between our system and turbulence theory,
we investigate an integration of energy spectrum, $K(t)=\int E(k,t)dk$, called the vortex energy in this paper. According to the theory of freely decaying turbulence~\cite{Batchelor1948,Comte-Bellot1966}, it obeys the power law $K(t)\propto  t^{-p}$ with $1\lesssim p \lesssim 1.2$.\par

To describe the decay law more precisely, we define the vortex energy as follows
\begin{equation}\label{vortex_energy}
K(t) \equiv \int^{1/\xi}_{1/L} E(k,t)dk
 =\frac{\xi^2}{\tau^2}\int^{l/\xi}_{l/L} \frac{\xi}{l}\tilde{E}(\tilde{k})d\tilde{k}.
\end{equation}
Here, lower and upper cutoffs are introduced for the integrations.
The former cutoff is not important and can be replaced by zero when $L$ is sufficiently larger than $l$.
The latter comes from the fact that there is no energy dissipation in our system and thus noise accumulates at high wavenumbers, which contrasts with the turbulence theory where the high wavenumber components are dissipated by viscosity.

To quantitatively evaluate our theory, we make further refinements.
A qualitative analysis for the free decaying turbulence gives $K(t)\propto E_0(t)k_0(t)$~\cite{Kida&Yanase1999}.
Here, $k_0$ is the wavenumber that gives the maximum value of the energy spectrum: $E_0=E(k_0)$.
This estimate corresponds to the energy spectrum having a sharp peak. On the other hand, in our system, the microscopic and macroscopic regimes are smoothly connected at $k_0=1/l$ and the peak is rounded. This difference produces a quantitative discrepancy.
To resolve this difference, we introduce an ansatz for the scaling function of the energy spectrum as
\begin{align} \label{ansatz}
    \tilde{E}(\tilde{k}) = \frac{2\tilde{E}_0\tilde{k}^{a}}{1+\tilde{k}^{a-b}}
\end{align}
with {$\tilde{E}_0 \equiv E_0 \tau^2/\xi^3$}.
This function takes the maximum value $\tilde{E}_0$ and is consistent with the asymptotic behavior of Eq.~\eqref{Asymp_scaling}.

Performing the integration yields
\begin{align}\label{Kt}
\frac{K}{\xi^2/\tau^2} =\frac{2\tilde{E}_0}{1+a}\frac{\xi}{l}\mathcal{F}
\end{align}
where we used 
$ \mathcal{F}=(l/\xi)^{1+a} \,_2F_1\left( 1,\frac{1+a}{a-b} ,1+\frac{1+a}{a-b} ;-(l/\xi)^{a-b}\right)
 -(l/L)^{1+a} \,_2F_1\left( 1,\frac{1+a}{a-b} ,1+\frac{1+a}{a-b} ;-(l/L)^{a-b}\right)$
with the Gauss hypergeometric function $_2F_1(\alpha,\beta,\gamma;z)=\sum_{n=0}^{\infty}\frac{(\alpha)_n(\beta)_n}{(\gamma)_n}\frac{z^n}{n!}$~\cite{Abramowitz1964}.
By assumption the peak should be independent of time, $\tilde{E}_0=5.46 \times 10^{12}$.
As is clear from Fig.~\ref{scaling}(b), the exponent in the macroscopic regime may be also independent of time.
This treatment is supported by the fact that the statistical behavior of that regime is firmly determined by the percolation criticality from the early stage~\cite{HT2018}.
The least squares method for the early stage determines the exponent as $a=2.35(4)$, satisfying the condition $-1<a\leq 4$ in~\cite{Kida1981}. The exponents are calculated in the region $0.1<\tilde{k}<0.5$ for the macroscopic regime and in $4<\tilde{k}<20$ for the microscopic regime~\cite{lsm}.

Since the slope of the logarithmic plot in the microscopic regime changes gradually with time from the early to late stage,
the exponent $b$ is determined based on a physical interpretation as follows.
It is important to note that the random arrangement or uncorrelated distribution of point vortices and vortex sheets gives $-1$ and $-2$ as the slope of the spectrum, respectively~\cite{Townsend1951}.
According to previous studies,
circular domain walls as quantum vortices are nucleated by quantum KHI at long domain walls,
which explains the anomalous behavior of the microscopic regime of the domain size distribution in the late stage.
In other words, the vorticity is originally distributed along long domain walls by forming randomly arranged vortex sheets in the early stage. In short, one would predict the slope to be closer to $-2$ in the early stage and gradually approaching $-1$ in the late stage.\par

This prediction partially explains the time evolution of slope in the microscopic regime in Fig.~\ref{fluid_energy}(a).
Also, as seen in Fig.~\ref{fig:1}, we typically found that only a few vortices exist between long domain walls in the late stage,
which can be thought of as making just an uncorrelated distribution of vortices. Based on the above considerations, we use $b=-1$ as the late-stage slope:
\begin{equation} \label{F_-1}
\mathcal{F}=\ln\left[1+\left(\frac{l}{\xi}\right)^{1+a}\right]-\ln\left[1+\left(\frac{l}{L}\right)^{1+a}\right]
\end{equation}
In the ideal situation $\xi \ll l \ll L$ for the dynamic scaling, we have
\begin{equation} \label{decay_asym}
\frac{K}{\xi^2/\tau^2} \sim 2\tilde{E}_0 \left(\frac{l}{\xi}\right)^{-1} \ln \frac{l}{\xi} \propto t^{-2/3}\ln~t,
\end{equation}
distinguishable from existing decay laws of freely decaying turbulence~\cite{Batchelor1948,Comte-Bellot1966,Kida1981}.

Figure~\ref{fluid_energy}(b) compares the numerical data plot of the vortex energy with the theoretical results based on Eqs.~\eqref{Kt} and ~\eqref{ansatz} with $b=-1$.
The numerical and theoretical results are in good quantitative agreement in the late stage.
At the end of the time evolution, there is a slight discrepancy between the numerical and theoretical plots.
This is thought to be due to the accumulation of energy of noise near the wavenumber corresponding to the spatial resolution of the numerical calculation.

{\it Summary and Discussion.}
This work succeeds in formulating the universality of the dynamics from the behavior of vorticity after phase separation while most of the previous work has focused on evaluating the structure factor of magnetization.
We found a dynamic scaling law of Eq.~\eqref{Scaling_function_Sv} with $\alpha=1$ for the vorticity structure factor in the non-equilibrium dynamics of phase-separating superfluid mixtures.
This law is supported by the fact that the decay of the vortex energy is quantitatively described by the ansatz [Eq.~\eqref{ansatz}] with $\tilde{E}_0=\mathrm{const}$.
Our results statistically implies that a few vortices exist in the region between long domain walls for $\xi \ll l \ll L$, causing $E(k)\propto k^{-1}$ at high wavenumbers.

Our theory is applicable to ferromagnetic superfluid of $^7$Li cold atoms under strong quadratic Zeeman shift.
Evaluating the slope $a$ in the macroscopic regime, which is associated with percolation criticality, remains an interesting topic for future work.
Although the statistical behavior at ultrahigh wavenumbers $k\xi \gg 1$ is thought to involve energy cascades due to phonons and ripplons, the numerical analysis of these cascades is not easy beyond the scope of this study.
Recent experiments reveal the power spectrum of ripple mode of a stable interface~\cite{Geng2024}, and this may open up the possibility to explore the physics of the ultra-high wavenumber regime in phase-separating superfluid mixtures.

\begin{acknowledgments}
H.T is supported by JSPS KAKENHI Grants No. JP18KK0391, and No. JP20H01842; and JST, PRESTO (Japan) Grant No. JPMJPR23O5.
\end{acknowledgments}

\appendix

\section{Numerical simulations}
The numerical simulation of the coupled Gross-Pitaevskii (GP) equations is computed on a two-dimensional lattice of square grids. The size and number of the numerical grids are set as $\Delta x/\xi=0.5$ and $L/\Delta x=4096$, respectively.
 After phase separation, the initial domain patterns with a characteristic length $l_0 \propto 1/\sqrt{g_{+-}/g-1}$ emerges at $\tau_0 \sim \frac{\tau}{g_{+-}/g-1}\ln\frac{\bar{n}}{2\delta^2}$ with $\tau=\hbar/\mu$ and a small amplitude $\delta=0.01$ of the initial seeds. Our numerical simulations were computed by the Crank-Nicolson method under periodic boundary conditions.

\section{Calculation of the characteristic length and the vorticity}
We use the mean distance between domain walls as the characteristic length defined by $l=L^2/R$ with the system size $L$ and the total length of the domain walls $R$. Then, $R$ is calculated by a collection of sides between neighboring grid points with $m=(n_+-n_-)/(n_++n_-)>0$ and $m < 0$. 
A domain is surrounded by a closed domain wall (or the system boundary and a open domain wall that ends at the boundary).

The dynamic exponent $z$ is estimated by using the least squares method (LSM), based on data for $t/\tau \geq 1000$. We also estimate the local-time exponent by applying LSM to equally spaced segments on a logarithmic time scale after averaged over 10 realizations.

The circulation is defined as
\begin{align}
\Gamma=\oint_C {\bm v} \cdot d{\bm l} = \int_S (\nabla \times {\bm v}) \cdot d{\bm S},
\end{align}
where we used the Stokes' theorem, $C$ is the closed loop on neighboring numerical grids, and $S$ is the mesh area inside the loop $C$. The velocity ${\bm v}$ is defined as Eq.~(1). The vorticity ${\bm \omega} = \nabla \times {\bm v}$ is computed by dividing the circulation $\Gamma$ by the area of mesh $\Delta S=(\Delta x)^2$.

\section{Formulation of universal functions}
According to the dynamic scaling hypothesis~\cite{Bray1994}, the vorticity structure factor is assumed to be rescaled by the characteristic length $l$ and can be described as a function of rescaled wavenumber $\tilde{k}=kl$. We therefore introduce the following dimensionless universal function of the vorticity structure factor with a scale exponent $\alpha$:
\begin{align}
    \tilde{S}_\mathrm{v}(\tilde{k}) \equiv \frac{S_\mathrm{v}(k,t)\, l(t)^{\alpha}}{\tau^{-2}\xi^{\alpha}}
\end{align}
with $\tau = \hbar / \mu$ and $\xi = \hbar / \sqrt{m\mu}$. Similarly, we define the universal function of the energy spectrum by the relation $E=L^2S_{\rm v}/(4\pi k)$ as
\begin{align}
    \tilde{E}(\tilde{k}) \equiv \frac{E(k,t)}{\xi^3 / \tau^2} = \frac{1}{4\pi} \left( \frac{L}{\xi} \right)^2 \frac{\tilde{S}_{\mathrm{v}}(\tilde{k})}{\tilde{k}} \left[ \frac{l(t)}{\xi} \right]^{1 - \alpha}.
\end{align}

\section{Estimation of fitting parameters}
The ansatz for universal function of the energy spectrum [Eq.~(9)] has three fitting parameters $\tilde{E}_0$, $a$ and $b$. 
The maximum values of the energy spectrum $\max(\tilde{E})$ show the time dependence through the entire time evolution [Fig.3(a)]. We determine $\tilde{E}_0=5.46(13)\times10^{12}$ as the average over the late stage where the dynamic scaling holds. The exponents of the universal function in each regime can be written as $d(\ln\tilde{E})/d(\ln\tilde{k})$. We estimate the exponent in the macroscopic regime as 
\begin{align}
    a=\left\langle\frac{d(\ln\tilde{E})}{d(\ln\tilde{k})}\right\rangle = 2.35(4).
\end{align}
Here, $\langle\cdots\rangle$ denotes the average over the early stage, calculated in the region $0.1<\tilde{k}<0.5$ and using LSM. 
We ignore the late stage to estimate the exponent $a$ because there are insufficient data points in this stage. The exponent in the microscopic regime is set as $b=-1$, based on a physical interpretation (see the main text). The calculation of the exponent in this regime at each time step is performed in the region $4<\tilde{k}<20$, using LSM. The scaling plots show that the scaling behaviors are established well in these regimes. The widths used to calculate the exponents in each regime are the same on a logarithmic scale.

\bibliography{cite}

\end{document}